\documentclass[11pt]{article}
\def\be{\begin{equation}}
\def\ee{\end{equation}}
\def\ba{\begin{eqnarray}}
\def\ea{\end{eqnarray}}

\def\ra{\rangle}

\def\h{\hskip 1cm}

\def\A1{A_{-1}}
\usepackage{epsfig}
\begin{document}
\begin{titlepage}
\vspace{4cm}

\begin{center}{\Large \bf New Phase Transitions in Optimal States for Memory Channels}\\
\vspace{2cm}\h Vahid Karimipour$ ^\dag$  \footnote{Corresponding
author:vahid@sharif.edu},\h Zohreh
Meghdadi\footnote{email:zohrehmeghdadi@physics.sharif.edu}$
^\dag$, \h Laleh Memarzadeh$
^\S$\footnote{email:laleh.memarzadeh@unicam.it
}, \\
\vspace{1cm} $ ^{\dag}$Department of Physics, Sharif University of
Technology,\\
 11155-9161, Tehran, Iran\\
\vspace{5 mm} $ ^\S$Dipartimento di Fisica, Universit`a di
Camerino, I-62032 Camerino, Italy.

\end{center}

\vskip 2cm

\begin{abstract}
We investigate the question of optimal input ensembles for memory
channels and construct a rather large class of Pauli channels with
correlated noise which can be studied analytically with regard to
the entanglement of their optimal input ensembles. In a more
detailed study of a subclass of these channels, the complete phase
diagram of the two-qubit channel, which shows three distinct phases
is obtained. While increasing the correlation generally changes the
optimal state from separable to maximally entangled states, this is
done via an intermediate region where both separable and maximally
entangled states are optimal. A more concrete model, based on random
rotations of the error operators which mimic the behavior of this
subclass of channels is also presented.
\end{abstract}
\vskip 2cm PACS Numbers: 03.67.HK, 05.40.Ca \hspace{.3in}
\end{titlepage}
\section{Introduction}\label{intro}

A basic question in quantum information theory \cite{Nil,
M,Holevo,Schumacher,Shor} is whether the use of entangled states
for encoding classical information can increase the rate of
information transmission though a channel or not. A proper
calculation of the so called Holevo capacity \cite{Holevo} of a
channel, representing by a Completely Positive Trace preserving
(CPT) map $\Phi$, requires the optimization of Holevo information
over ensemble of input states when we encode information into
arbitrary long strings of quantum states (more precisely states
in the tensor product of the Hilbert space of one state) and
carrying out the limiting procedure $C:=\lim_{n\rightarrow\infty}
C_n$, where
\begin{equation}\label{}
    C_n:=\frac{1}{n}Sup_{\varepsilon}\chi_n(\varepsilon)
\end{equation}
is the capacity of the channel, when we send strings of $n$
quantum states into the channel. Here
$\varepsilon:=\{p_i,\rho_i\}$ is the ensemble of input states,
\begin{equation}\label{holevo}
    \chi_n(\varepsilon):= S(\sum_ip_i\Phi(\rho_i))-\sum_{i}p_iS(\Phi(\rho_i))
\end{equation}
is the Holevo information of the ensemble and $S(\rho)\equiv
-tr(\rho \log \rho)$ is the von Neumann entropy of a state $\rho$.
To find the capacity of a given channel we should find the
ensemble which maximizes this quantity and we call it the optimal
ensemble of input states. Then the properties of this ensemble
can be studied and ask whether this ensemble
includes entangled states or not.\\
\\
The importance of this question stems from the fact that
entanglement is a vital quantum mechanical resource in many tasks
in quantum information processing, however it is a held belief
that entanglement is so fragile in the presence of noise. So it
would be interesting if one can show by using this property higher
rate of data transmission through noisy channels is achievable.
The difficulty in answering this question is not only because of
the optimization of Holevo information over multi-parameter space
but also due to the fact that no concrete classification of
multi-particle entangled states exist.\\
\\
A much simpler problem is to calculate $C_2$, rather than $C_n$,
which is equivalent to refresh the channel after each two uses,
and see whether entangled states can enhance Holevo information
or not. This problem has been tackled by many authors
\cite{M1,K1,K2,M2,Matsu,Auden,M3,C,KM1,karpov,KM2} and for a kind
of correlated channel it has been shown that entanglement can
enhance the Holevo information, if the correlation is above a
certain threshold value.
This correlated channel was first introduced in \cite{M2} as
follows
\begin{equation}\label{MachDef}
    \Phi(\rho) = \sum_{i,j=0}^3 P_{ij} \sigma_{i}\otimes
    \sigma_{j} \rho \sigma_{i}\otimes \sigma_j,
\end{equation}
where $\sigma_i$, $i=0, 1, 2, 3$ represent Pauli operators $I,
\sigma_x, \sigma_y$ , $\sigma_z$ and $P_{ij}$, the probabilities
of errors $\sigma_i\otimes \sigma_j$ are correlated in a special
way, namely
\begin{equation}\label{pij1}
    P_{ij}=(1-\mu)p_ip_j+\mu\delta_{ij}p_j.
\end{equation}
The parameter $\mu\in[0,1]$, called memory factor signifies the
amount of correlation in the noise of the channel. For $\mu=0$,
the errors on the two consecutive qubits are completely
un-correlated, while for $\mu=1$, the two errors are exactly the
same. The idea behind this, is that when the channel relaxation
time is much less than the time interval between the passage of
the two qubits, the errors will be the same. What was observed in
\cite{M2} and in subsequent works \cite{M3,KM1,KM2} was that when
$\mu$ passes a critical value $\mu_c$, the optimal input state
jumps suddenly from product states to maximally entangled
states.\\
\\
It is important to note that although the sharp transition of
optimal ensemble of input states from product to maximally
entangled states is interesting and cause non-analytical behavior
of the channel, it is not obvious that the same happens for all
kinds of correlated channels.
%
the correlation (\ref{pij1}) is one out of many different forms
of correlations that one can envisage for a correlated noisy
channel. Apart from the mathematical possibility of defining many
other forms of correlations, one can also argue on physical
grounds, in favor of other forms of correlation. A fully
correlated channel which exerts a noise operator on the first
qubit, need not exerts the same error operator on the second
qubit, as in (\ref{pij1}). In fact it is natural to expects that
on exerting the first error, the state of the environment will
change and depending on this new state, it will exert new errors
with conditional probabilities on the second qubit even if the
time interval between two consecutive uses of the channel be
small.\\
\\
In this paper we try to shed light on these issues and to provide
a basis for studying more examples and extend the study of
correlated noisy channels by considering more general forms of
correlations. Our study will be along the line of references
\cite{M2,M3,KM1,KM2}, that is we do not consider a specific model
of environment, rather we take the abstract definition of the
channel as a CPT linear map, defined by its Kraus decomposition
\cite{K}. We start in section 2 with the most general form of the
action of a correlated Pauli channel on two qubits, where by two
plausible requirements, we will restrict the parameters so that
the members of this class can be studied by analytical means.
Then we focus on one particular subclass and study in detail the
optimal ensemble which maximizes its Holevo information. In this
same section we propose a general definition of the correlation
parameter, and show that when the correlation of the noise in
this channel increases, the optimal ensemble changes from a
product ensemble to a maximally entangled one. The phase diagram
of the model also shows other interesting transitions not already
observed in other works \cite{M2,M3,KM1,KM2}. In fact the phase
diagram contains three distinct phases, a product one and two
maximally entangled ones, which differ with each other by a
relative phase.  We also observe that when we consider the
correlation parameter and move through this phase diagram,  the
optimal ensemble changes from product to maximally entangled,
however this transition is mediated by a region of correlation,
where both maximally entangled and separable ensembles are
optimal. Finally we construct a rather concrete model for this
particular type of correlation. The paper concludes with a
discussion.

\section{The correlated action of a Pauli channel on two qubits }
The general action of a Pauli channel on two qubits is defined by
the following CPT map
\begin{equation}\label{canal2}
    \Phi(\rho) = \sum_{ij} P_{ij} \sigma_i\otimes \sigma_j \rho \sigma_i
    \otimes \sigma_j,
\end{equation}
where $P_{ij}$ is the probability of the errors $\sigma_j$ and
$\sigma_i$ on the first and second qubits entering the channel
respectively. The number of independent error probabilities are
15 due to normalization $\sum_{i,j}P_{ij}=1$.\\
\\
In order to see how much correlated the noise on the two
consecutive qubits are, we form the marginal error probabilities
$p^{(1)}$ and $p^{(2)}$ and evaluate the following distance
between the two probability distributions, $p_{ij}$ and
$p^{(1)}_ip^{(2)}_j$, denoted as ${ \cal C}$

\begin{equation}\label{distance}
    {\cal C}:=\frac{1}{2}\sum_{i,j=0}^3
    |P_{ij}-p^{(1)}_ip^{(2)}_j|
\end{equation}
where $p^{(1)}_i:=\sum_j P_{ij}$ and $p^{(2)}_i=\sum_j P_{ji}.$
For uncorrelated noise we will have ${\cal C}=0$ and for a fully
correlated noise ${\cal C}$ will attain its maximum value. For the
probability distribution (\ref{pij1}), ${\cal C}$ turns out to be
${\cal C}=\mu\sum_{i=0}^3 p_i(1-p_i),$ hence for fixed error
probabilities, the correlations indeed
increase with $\mu$. \\
\\
There are two difficulties in studying such a channel for
obtaining its optimal input ensemble and understanding how it
depends on the correlation of the noise. The first one is that
the probability distribution $P_{ij}$ can be correlated in many
different ways and picking out a single parameter and designate it
as the memory of the channel, is just one possibility. The second
problem is that the manifold of input states has itself 6 real
parameters which means that the optimization task should be done
over a 6 parameter space and thus analytical treatment of such a
channel is almost impossible. To overcome these problems, we
impose the following symmetry on the channel,
\begin{equation}\label{symmetry}
     \Phi(\sigma_3\otimes \sigma_3\rho\sigma_3\otimes
    \sigma_3)= \Phi(\rho).
\end{equation}
\\
This is the symmetry which has been considered in \cite{M3} for
making the model amenable to analytical treatment. Demanding this
symmetry reduces the number of parameters in $P_{ij}$ to 7
parameters:
\begin{equation}
    P = \left(\begin{array}{cccc}
    p & t & u & s \\
    v & q & r & w\\
    w & r & q & v\\
    s & u & t & p
    \end{array}\right),
\end{equation}
with normalization relation between the parameters. However it is
more plausible to assume that the marginal error probabilities on
the first and the second qubits be equal, that is
\begin{equation}\label{marginal}
    p^{(1)}_i=p^{(2)}_i\h \forall \ i,
\end{equation}
Here we are assuming that the errors on a sequence of qubits
should be the same, regardless of how we enumerate the qubits of
the sequence. What really matters is that any two consecutive
uses of the channel are correlated.\\
\\
This assumptions in addition to the constrained composed by the
symmetry in (\ref{symmetry} reduce the number of parameters from
to 6 and the final form of the matrix of probabilities $P$ with
elements can be parameterized as follows:
\begin{equation}\label{pij2}
    P = \left(\begin{array}{cccc} p & \frac{\eta+\xi}{4} & \frac{\eta-\xi}{4} & s \\
    \frac{\eta+\gamma}{4} & q & r & \frac{\eta-\gamma}{4}\\
    \frac{\eta-\gamma}{4} & r & q & \frac{\eta+\gamma}{4}\\
    s & \frac{\eta-\xi}{4} & \frac{\eta+\xi}{4} &
    p\end{array}\right),
\end{equation}
where $p+q+r+\eta+s=\frac{1}{2}.$\\
The advantage of demanding the symmetry in (\ref{symmetry}) in
not only in reducing the parameters of $P$ but also in
reducing the parameters of the general input states.\\
\\
Following \cite{M3}, we form the optimal ensemble by finding a
state $\rho^*$ which minimizes the output entropy and hence
minimizing the second term in the right hand side of
(\ref{holevo}). The input ensemble is then formed as a uniform
distribution of the states ${\cal E} =
\{\rho_{ij}:=(\sigma_i\otimes \sigma_j)\rho^* (\sigma_i\otimes
\sigma_j)\}$. The reason is that the first term of the Holevo
quantity is maximized by this choise, i.e.
\begin{eqnarray}\label{firstterm}
    S(\sum_{ij}p_{ij} \Phi(\rho_{ij}))&=& S(\frac{1}{16}\sum_{ij}\sigma_i\otimes \sigma_j\Phi(\rho^*)\sigma_i\otimes
    \sigma_j)\cr &=& S(\frac{1}{4}I)=2,
\end{eqnarray}
where in the first line we have used the covariance property of
the channel:
\begin{equation}\label{fullcovariance}
\Phi(\sigma_i\otimes \sigma_j \rho \sigma_i\otimes \sigma_j) =
(\sigma_i\otimes \sigma_j)\Phi(\rho)(\sigma_i\otimes \sigma_j).
\end{equation}
and in the second line the Shur's first lemma for irreducible
representations of the Pauli group.\\
\\
Therefore finding the optimum ensemble of input states reduces to
finding a single input state which minimizes the output entropy
and we call it optimum input state. Regarding the convexity of
entropy we deduce that we should search for the optimal input
state among pure states which in general has 6 parameters.
However, we are able to restrict the form of the input states to
the simple states which are invariant under the above symmetry in
(\ref{symmetry})
\begin{equation}\label{state}
    |\psi\ra = \cos \theta |00\ra + \sin \theta e^{i\phi}|11\ra,
\end{equation}
It is easy to find the output state of this channel with the above
form of error probabilities. A straightforward calculation shows
that the output state, in the computational basis $\{|00\ra,
|01\ra, |10\ra, |11\ra\}$ will be
\begin{equation}\label{outputstate}
    \Phi(\rho) = \left(\begin{array}{cccc} \varepsilon_{00} & 0 & 0 & \varepsilon_{03}
    \\ 0 & \varepsilon_{11} & \varepsilon_{12} & 0
    \\ 0 & \varepsilon_{12}^*& \varepsilon_{22}& 0
     \\ \varepsilon_{03}^* & 0 &  0 & \varepsilon_{33}\end{array}\right)
\end{equation}
where
\begin{eqnarray}\label{outputeig}
\varepsilon_{00}&=&2(p+s) \cos^2 (\theta)+2(q+r)\sin^2 (\theta)\cr
\varepsilon_{33}&=&2(p+s) \sin^2 (\theta)+2(q+r)\cos^2 (\theta)\cr
\varepsilon_{03}&=&\sin(2\theta)[(p-s)e^{-i\phi}+(q-r)e^{i\phi}]\cr
\varepsilon_{12}&=&\frac{1}{2}\sin(2\theta)[\xi e^{-i\phi}+\gamma
e^{i\phi}]\cr \varepsilon_{11}&=&\varepsilon_{22}  = \eta.
 \end{eqnarray}
Due to the block diagonal structure of this matrix, its
eigenvalues can be calculated in closed form and hence a complete
specification of the optimal input states can be made for this
general class of correlated Pauli channels.

\subsection{Detailed study of a subclass and its phase diagram}

As an interesting and simple example, we consider a channel where
only the parameters $p$, $q$ and $r$ are non-vanishing, that is
matrix of probabilities has the following form
\begin{equation}\label{pij}
    P=\left(\begin{array}{cccc} p & 0 & 0 & 0 \\
0 & q & r &0 \\ 0 & r & q & 0 \\ 0 &0 & 0 & p \\
    \end{array}\right),
\end{equation}
where due to normalization $p+q+r = \frac{1}{2}.$ The action of
the Pauli channel on two consecutive qubits will then be:
\begin{eqnarray}\label{canal2simple}
     \Phi(\rho) &=& p \rho + p \sigma_3\otimes \sigma_3 \rho \sigma_3\otimes \sigma_3 + q \sigma_1\otimes \sigma_1 \rho \sigma_1\otimes \sigma_1 +
     q \sigma_2\otimes \sigma_2 \rho \sigma_2\otimes \sigma_2 \cr & + &  r \sigma_1\otimes \sigma_2 \rho \sigma_1\otimes \sigma_2
    + r \sigma_2\otimes \sigma_1 \rho \sigma_2\otimes \sigma_1.
    \end{eqnarray}
The correlation parameter of this channel is found to be
\begin{equation}\label{distanceExample}
    {\cal C}:=3p-4p^2+ |(q+r)^2-q|+|(q+r)^2-r|.
\end{equation}
Using (\ref{outputeig}), the eigenvalues of the output density
matrix are
\begin{equation}\label{eigenvalues2}
    \lambda_{1,2}=0,\h \lambda_{3,4}=\frac{1}{2}\left(1\pm
    \sqrt{1-16[p(q+r) + Y\sin^2 (2\theta)]}\right),
\end{equation}
where
\begin{equation}\label{Y}
Y:=q(r-p)\cos^2\phi +r(q-p)\sin^2 \phi.
\end{equation}
For minimization of output entropy which is
$S(\Phi(\rho))=-\lambda_1 \log \lambda_1 - \lambda_2\log
\lambda_2$, we should maximize the difference between $\lambda_1$
and $\lambda_2$. This is achieved by minimizing $Y\sin^2 2\theta$.
Therefore if $Y\leq 0 $, we should choose $\theta = \frac{\pi}{4}$
and if $Y\geq 0 $, we should choose $\theta = 0$. Therefore the
line $Y=0$ determines the boundary of the maximally entangled
($\theta =\frac{\pi}{4}$) and the product ($\theta=0$) optimal
states. This line is determined by the equations $q(r-p)=0$ and
$r(q-p)=0$. In the $(q,r)$ plane these two lines are specified by
the equations
$2q+r = \frac{1}{2}$ and $2r+q=\frac{1}{2}$.\\

\begin{figure}[t]
 \centering
  \includegraphics[width=7cm,height=7cm,angle=0]{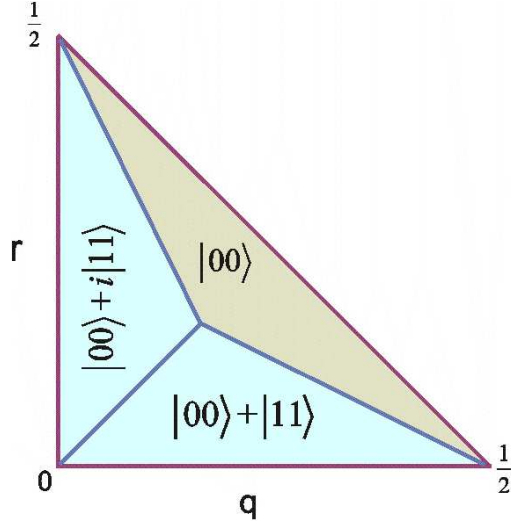}
 \caption{Color Online. The phase diagram of model (14).
 In each phase the minimum output entropy state (un-normalized) is specified. The triple point is $(q,p)=(\frac{1}{6},\frac{1}{6}).$ }
\label{BasicPhaseDiagramColor}
\end{figure}

When $\theta=0$, the value of $Y$ is immaterial, however when
$\theta=\frac{\pi}{4}$, then the value of $Y$ should be minimized
again. In the maximally entangled region, where
$\theta=\frac{\pi}{4}$, we should still minimize $Y$. For
$q(r-p)\geq r(q-p)$, $Y$ takes its minimum at $\phi=0$, and for
$q(r-p)< r(q-p)$ it will take its minimum at $\phi=\frac{\pi}{2}$.
Thus the line $q(r-p)=r(q-p)$ ($q=r$) will separate two types of
optimal maximally entangled states from each other. The phase
diagram is shown in figure (\ref{BasicPhaseDiagramColor}).

\begin{figure}[t]
 \centering
  \includegraphics[width=9.7cm,height=8.5cm,angle=0]{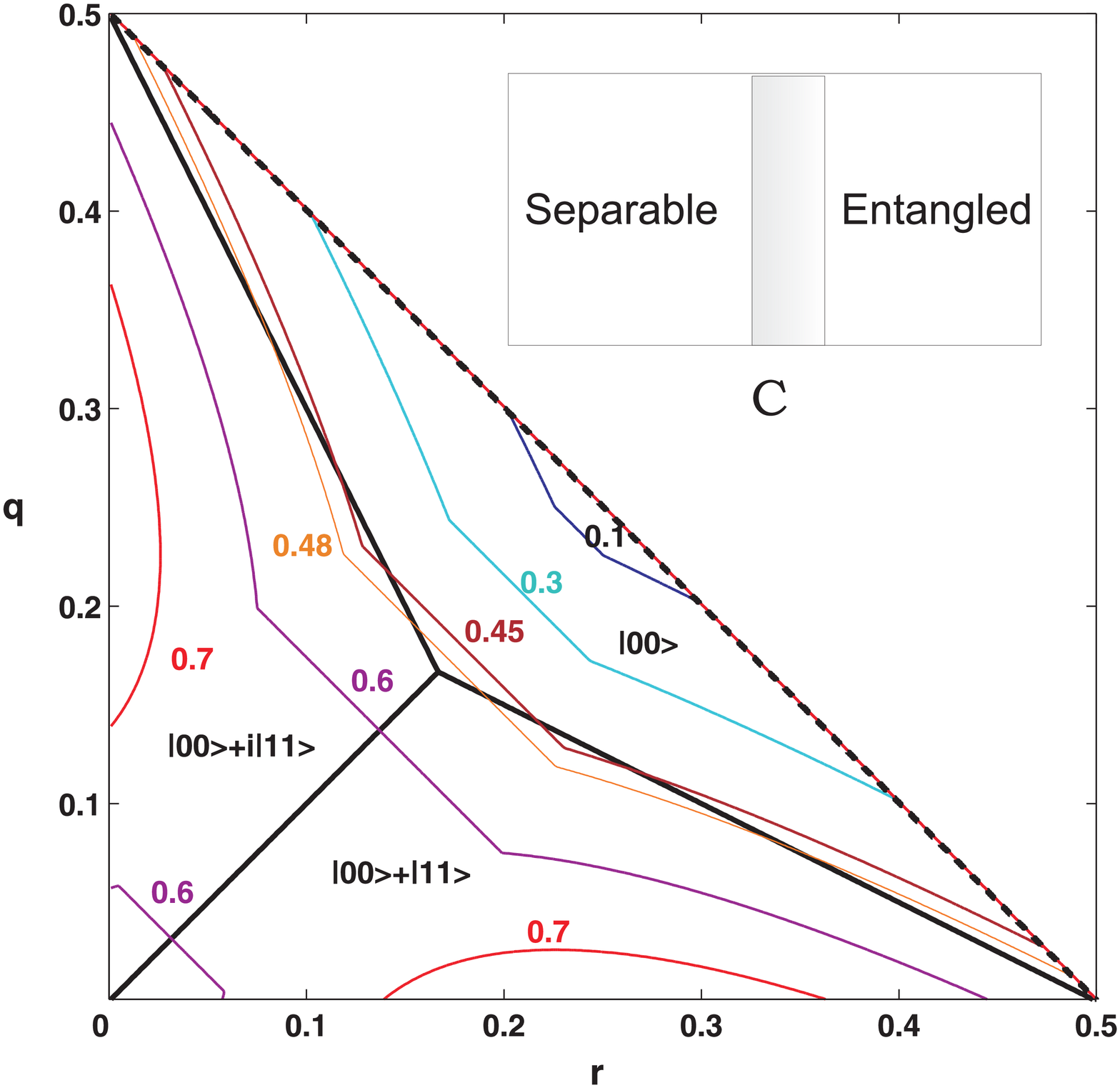}
 \caption{Color Online. The phase diagram of model (14). In each phase the minimum output entropy state (un-normalized) is specified. The contours of constant
 correlations are also shown. As the correlation ${\cal C}$ increases, the optimal ensemble changes from separable to maximally entangled one.
 For $\approx 0.43 < {\cal C}< \approx 0.5 $, both types of states are optimal, the grey region shown in the inset. }
\label{final3}
\end{figure}

It is seen from the phase diagram (\ref{final3}) that as the
correlation parameter increases, the optimal input ensemble changes
from product to maximally entangled. There are however two
remarkable features in this diagram not encountered in
previous studies.\\

First we see that depending on the values of the channel parameters,
two different types of maximally entangled states, namely
$\frac{1}{\sqrt{2}}(|00\ra+|11\ra)$ and
$\frac{1}{\sqrt{2}}(|00\ra+i|11\ra)$ are optimal. Although these two
types of states, are transformed to each other by a local operator
$I\otimes \left(\begin{array}{cc} 1 & 0
\\ 0 & i\end{array}\right)$, since the channel is not covariant under this
local operator, they should be considered different as far as
optimality of the encoding is concerned, although they are
equivalent as far as their entanglement properties are concerned
\cite{entClasses}.\\

Second, when we draw the contours of constant correlations in this
phase diagram, we observed that there is a region of correlation,
for which both separable and maximally entangled states are optimal,
depending on the values of the parameters $q$ and $r$, figure
(\ref{final3}). We say that the two phases coexist, a property which
is reminiscent of first-order transitions. We should stress that if
one draws the phase diagram of the model in \cite{M3}, not in terms
of the parameter $\mu$, but in terms of the correlation parameter
$C$, one will again see such coexistence region. Therefore and
specially with regard to recent studies in relating transitions in
channel capacity to the critical transitions in their environment
\cite{PlenioVirmani, Giovannetti}, it is an interesting issue to see
if transitions in
the channel should be characterized as first or second order.\\

\subsection{A concrete and intuitive model of correlation}

In this section we construct a particular model for correlated
noise in Pauli channels, which will reproduce the above example of
correlation in a natural way. Consider a noisy Pauli channel
acting on a qubit, defined as
\begin{equation}\label{Pauli1}
    \Phi(\rho) = \sum_{i=0}^3 p_i \sigma_i \rho \sigma_i,
    \end{equation}
where $p_i$ is the probability of error $\sigma_i$ ($\sigma_0=I$)
and $\sum_{i=0}^3 p_i = 1$. When the first qubit passes through
the channel, and an error operator $\sigma_i$ acts on it, we
assume that the state of the channel changes randomly and
therefore on the second qubit, it exerts not the same error or a
fixed error for that matter, but a random rotation of the
$\sigma_i$ operator, in the form
\begin{equation}\label{sigmatilde}
    \tilde{\sigma}_i := U_{{\bf n},\theta}\sigma_i U^{\dagger}_{{\bf
    n},\theta},
\end{equation}
where $U_{{\bf n},\theta}$ is a  random rotation around the axis
${\bf n}$ with angle $\theta$. Thus $\tilde{\sigma}$ has the
effect of the first error operator and also the random change in
the environment. This randomness in contrast to a deterministic
change in the environment state is physically plausible in view
of the macroscopic nature of the environment.\\

Therefore the action of the channel on two consecutive qubits may
be written as follows
\begin{equation}\label{Pauli1}
    \Phi(\rho) = p_0 \rho + \sum_{i=1}^3 p_i \tilde{\sigma}_i\otimes \sigma_i\ \rho \ \tilde{\sigma}_i\otimes \sigma_i
    .
\end{equation}
Since the rotations are random, the complete definition of the
channel will be given by integrating over the above action with a
suitable probability distribution over random rotations. Thus the
final definition will be
\begin{equation}\label{SigmaModel}
\Phi_{\sigma}(\rho) = \int d{\hat {\bf n}}d\theta P({\bf n},
\theta)
 \Phi(\rho).
\end{equation}
Clearly one can add more parameters to the above model, for
example by taking different rotations to be along different axes.
For simplicity, let us restrict ourself to a simple example in
which the direction of all rotations are fixed in the $z$- axis
and only the angle of rotation is random. Also to ensure the
symmetry (\ref{symmetry}){ we take $p_0=p_3,$ and $ p_1=p_2$,
where $p_0+p_1=\frac{1}{2}$.  We take the probability
distribution to be a Gaussian with mean value $\theta_0$ and
variance $\sigma$. Hence the channel will be defined by

\begin{equation}\label{SigmaModel}
\Phi_{\sigma}(\rho) = \int \frac{d\theta}{\sigma\sqrt{2\pi}}
e^{-\frac{(\theta-\theta_0)^2}{2\sigma^2}} \Phi(\rho),
\end{equation}
where in this case $\tilde{\sigma}_i=e^{-\frac{i}{2}\theta
\sigma_z}\sigma_i e^{\frac{i}{2}\theta \sigma_z}$. One can say
that parameter $\sigma$ is related to the memory of the channel.
When $\sigma=0$, the channel has full memory and it will exert a
definite error operator (exactly the same error in the case
$\theta_0=0$) on the second qubit depending on the operator which
it has exerted on the first. However for a non-zero small value
of $\sigma$, the channel exerts errors on the second qubit which
are close to the errors on the first qubit. As $\sigma$ increases
further the memory is lost further and the channel will exert
errors
from a larger neighborhood of the errors on the first qubit.\\
\\
A remark is in order about the Gaussian distribution. The rotation
operators are periodic which restrict the range of integration of
$\theta$ to $[0,2\pi]$. However this makes the subsequent formulas
unduly cumbersome without adding much to the physics. Instead we
can assume the variance $\sigma$ to be sufficiently less than
$2\pi$ so that we can safely extend the range of integration of
$\theta$ to $(-\infty, \infty)$ and use the simple results of
Gaussian
integration. \\

After rearranging and doing the integrals, one finds that this
channel has the form (\ref{canal2simple}) with the parameters as
given below
\begin{eqnarray}
  p &=& p_0,\\
  q &=& p_1 (\frac{1+ e^{-2\sigma^2}}{2}),\label{1} \\
  r &=& p_1 (\frac{1- e^{-2\sigma^2}}{2}), \label{2}
\end{eqnarray}
The two independent parameters of this channel can be taken to be
$p_1$ and $\sigma$. In terms of these parameters the phase diagram
is shown in figure (\ref{sigmaphase}). Since we have always $r<q$,
the region with $\frac{1}{\sqrt{2}}(|00\ra+i|11\ra)$ optimal state
is not covered in this new phase diagram. The line which separates
the product phase from the maximally entangled phase is now given by
(\ref{sigma}). Inserting the values of $q$ and $r$ from (\ref{1})
and (\ref{2}) in the relation $2r+q=\frac{1}{2}$ and simplifying we
obtain

\begin{equation}\label{sigma}
    e^{-2\sigma^2}=3-\frac{1}{p_1}.
\end{equation}
The phase diagram in figure(\ref{final3}) is re-drawn in terms of
the new parameters in figure (\ref{sigmaphase}).

\begin{figure}[t]
 \centering
   \includegraphics[width=8.5cm,height=6cm,angle=0]{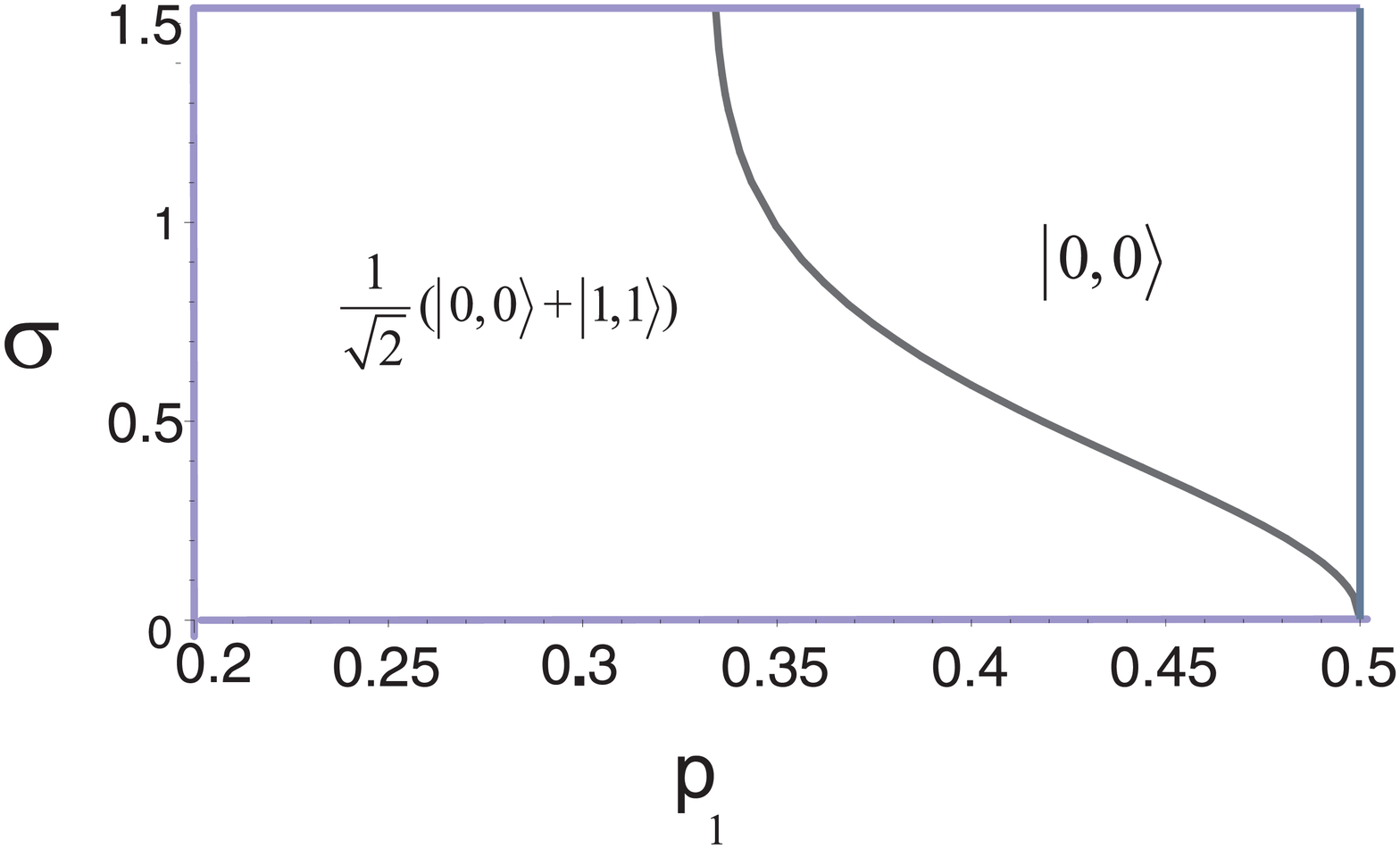}
   \caption{Color Online. The phase diagram of model (24). In each phase the minimum output entropy state is specified. }
   \label{sigmaphase}
\end{figure}
It is seen that depending on the value of $p_1$, the optimal
ensemble changes from separable to maximally entangled phase, when
the memory passes a certain threshold (note that here a lower
value of $\sigma$ means a larger value of memory). Also there are
values of $p_1$, where the optimal ensemble is always a maximally
entangled one, no matter how weak the memory is. This is related
to the fact that for no value of the parameter $\sigma$, this
channel is a product channel.\\
\\
As stressed in \cite{KM2}, the effect of memory on the type of
optimal ensemble input and ultimately on the capacity of a
channel, can be decided only when one considers the action of the
channel on arbitrary long sequences of (entangled) qubits. With
the type of memory introduced in \cite{M2}, such extension is
very difficult to pursue analytically and one can consider very
limited class of states. However with the type of memory
introduced above, we think that such an extension is indeed more
tractable by analytical means. For example one can consider the
following type of correlated action of a Pauli channel on strings
of $n$ qubits:

\begin{equation}\label{nqubit}
    \Phi(\rho) = \int dU_1 dU_2 \cdots dU_{n-1}\sum_{i=0}^3 p_i \Phi^{(n)}_i (\rho)
\end{equation}
where
\begin{equation}\label{Ei}
    \Phi^{(n)}_i(\rho) = [\sigma_i^{(n-1)}\otimes \cdots
    \sigma_i^{(1)}\otimes\sigma_i] \rho [\sigma_i^{(n-1)}\otimes \cdots
    \sigma_i^{(1)}\otimes\sigma_i]^{\dagger},
\end{equation}
and $\sigma_i^{(k)}=U_k \sigma_i U_k^{\dagger}$ in which $U_k$ is a
random unitary operator. Taking the random unitaries from a
distribution, one may relate the memory of the channel in a
qualitative sense to the properties of the distribution as we did
above.

\section{Discussion}

We have considered a large class of two-qubit correlated Pauli
channels for which the entropy of the output state can be determined
in closed analytical form. These channels are covariant under the
action of the two-qubit Pauli Group and have the symmetry
$\Phi(\rho)=\Phi(\sigma_z\otimes \sigma_z\rho \sigma_z\otimes
\sigma_z)$. For a subclass of these channels we have explicitly
determined the optimal input ensemble which maximizes the Holevo
capacity and have determined the exact phase diagram, showing
different regions where separable or maximally entangled states are
optimal. There are two new features of this phase diagram. First,
there are three phases separable and two different types of
maximally entangled states which are optimal. Second if we use a
precise definition of correlation, then as the correlation increases
the type of optimal state generally changes from separable to
maximally entangled states, however this is done through an
intermediate region where both separable and maximally entangled
states are optimal.  Put differently, if we think of correlation as
a control parameter,  the transition in our model is reminiscent of
first order transition where there are regions of coexistence of the
two phases. An interesting  question is whether such transition
occurs when the channel acts upon a string of qubits, not only two
consecutive qubits \cite{KM2}. This is a question which should be
addressed if we are to judge definitely wether or not encoding of
classical information enhances the classical capacity of quantum
channels. Unfortunately settling
this question seems to be very difficult.\\

Finally one would like to see if the transition in the quantum
capacity of quantum channels can be related to the well-studied
subject of critical phenomena. In this direction a formalism has
been developed in \cite{PlenioVirmani,Giovannetti} where models of
correlated noise are constructed by taking  a many body system as
the environment. The qubits which pass through the channel interact
with this many body environment and the correlations in the many
body state gives rise to correlations in the noise in the channel.
The question of whether the transitions in channel capacity are of
first order or not certainly have relevance for such studies.\\


{\textbf{Acknowledgements}} We would like to thank S. Alipour,  R.
Annabestani, S. Baghbanzadeh,  M. R. Koochakie, and A. Mani, for
valuable discussions. L.M. acknowledges financial support from the
EU project CORNER, number FP7-ICT-213681.

{}

\end{document}